\documentstyle[12pt,world_sci]{article}

\title{Minimal Quark-Lepton Symmetry Model \\
  and Possible Limits on Z'-mass from TRISTAN and LEP200}

\author{A.D.~Smirnov\thanks{E-mail: phystheo@univ.yars.free.net}\\
{\small\it Division of Theoretical Physics, Department of Physics,}\\
{\small\it Yaroslavl State University, Sovietskaya 14,}\\
{\small\it 150000 Yaroslavl, Russia.}}

\date{}

\begin{document}

\begin{flushright}
{\normalsize Yaroslavl State University\\
             Preprint YARU-HE-94/06\\
             hep-ph/9409448} \\[3cm]
\end{flushright}

\maketitle

\begin{abstract}

A minimal extension of the Standard Model containing the
four-color quark-lepton symmetry is discussed.
Some features of an additional $Z'$-boson originated from this
symmetry are investigated and the limits on $m_{Z'}$ from the
current TRISTAN data and the possible limits on $m_{Z'}$ from
the future LEP200 collider are obtained.

\end{abstract}

\vglue 3cm

\begin{center}
{\it Talk given at the VIII International Seminar "Quarks-94",\\ Vladimir,
Russia, May 11-18, 1994}
\end{center}

\newpage

The search for a new physics beyond the Standard Model (SM) is now
one of the main problems of the high energy physics.
It seems that the $SU_L(2) \times U(1) \times SU_c(3)$- symmetry
of the SM is only the first stage in the hierarchy of possible
symmetries of fundamental interactions. To search for the next such
stage it seems reasonable to investigate various minimal extensions
of the SM by adding to it some additional symmetries such as $SU_R(2)$-
symmetry, supersymmetry, etc.
One of such symmetries possibly existing
in nature and being worthy of the detailed investigation now is the
four-color quark-lepton symmetry regarding the lepton number as the
fourth color \cite{PSm}.

In this work we discuss the minimal quark-lepton symmetry model
of the unification of the strong and electroweak interactions (MQLS-model)
which is the minimal extension of the SM containing the four-color
quark-lepton symmetry and investigate the limits on the mass of $Z'$-
boson originated from this symmerty using the current TRISTAN data and the
possible limits on $m_{Z'}$ which can be achieved at LEP200.

The model to be discussed here is based on the $SU_V(4) \times SU_L(2)
\times U_R(1)$-group as the minimal group containing the four-color
symmetry of quarks and leptons. In this model the quarks $Q_{p a \alpha}
= \psi_{p a \alpha}$,
$a$ = 1, 2, $\alpha$ = 1, 2, 3 and the corresponding leptons $\ell_{p a}
= \psi_{p a 4}$ in the generation of the number $p = 1, \, 2, \, 3, \, \ldots$
form the four-color
fundamental quartet $\psi_{p a A}$, $A$ = 1, 2, 3, 4 of the $SU_V(4)$-group.
Under the $SU_L(2) \times U_R(1)$-group
the left fermions are the doublets with $Y_L$ = 0 and the
right fermions are the singlets with $Y_R$ = $\pm$ 1 for the ``up''
($a$ = 1) and ``down'' ($a$ = 2) fermions respectively.
For three generations the basic up- and down-fermion $SU_V(4)$-quartets are

\begin{displaymath}
{\psi'}_{p 1 A} \; : \quad
\left ( \begin{array}{c} {u'}_\alpha \\ {\nu'}_e \end{array} \right ) , \;
\left ( \begin{array}{c} {c'}_\alpha \\ {\nu'}_\mu \end{array} \right ) , \;
\left ( \begin{array}{c} {t'}_\alpha \\ {\nu'}_\tau \end{array} \right ) , \;
\cdots
\end{displaymath}

\begin{displaymath}
{\psi'}_{p 2 A} \; : \quad
\left ( \begin{array}{c} {d'}_\alpha \\ {e^-}' \end{array} \right ) , \;
\left ( \begin{array}{c} {s'}_\alpha \\ {\mu^-}' \end{array} \right ) , \;
\left ( \begin{array}{c} {b'}_\alpha \\ {\tau^-}' \end{array} \right ) , \;
\cdots
\end{displaymath}

\noindent where the basic quark and lepton fields
${Q'}^{L,R}_{p a \alpha}$,
${\ell'}^{L,R}_{p a}$
can be written, in general, as superpositions

\begin{eqnarray}
{Q'}^{L,R}_{p a \alpha} = \sum_q \left ( A^{L,R}_{Q_a} \right )_{p q}
Q^{L,R}_{q a \alpha} , \nonumber \\
{\ell'}^{L,R}_{p a} = \sum_q \left ( A^{L,R}_{\ell_a} \right )_{p q}
\ell^{L,R}_{q a}  \nonumber
\end{eqnarray}

\noindent of mass eigenstates
$Q^{L,R}_{q a \alpha}$, $\ell^{L,R}_{q a}$. Here
$A^{L,R}_{Q_a}$ and $A^{L,R}_{\ell_a}$ are unitary matrices.

The electric charges of quarks and leptons are related to the
generators of the group by

\begin{displaymath}
Q^{L,R} \, = \, \sqrt{\frac{2}{3}} \, t_{15}^{L,R} \, + \,
\frac{\tau_3^L}{2} \, + \, \frac{Y^R}{2} \, ,
\end{displaymath}

\noindent where $t_{15}$, $\tau_3/2$ are the corresponding generators,
$\tau_3$ is the Pauli matrix.

According to the structure of the group the gauge sector consists
of 19 fields $A_\mu^i$, $i = 1, 2, \ldots, 15$; $W_\mu^k$,
$k = 1, 2, 3$ and $B_\mu$. The first eight of them are the gluons
$G_\mu^j$ = $A_\mu^j$, $j = 1, 2, \ldots, 8$; the next six fields
form the triplets of the leptoquarks $V_{\alpha \mu}^\pm$, $\alpha = 1, 2, 3$
with the electric charge $Q_V = \pm 2/3$; $W^1_\mu$, $W^2_\mu$
form the $W^\pm$-bosons in a usual way and the remained fields
$A^{15}_\mu$, $W^3_\mu$, $B_\mu$ form the photon, the Z-boson
and an extra Z'-boson.

The electromagnetic field $A_\mu$ is related to $A^{15}_\mu$, $W^3_\mu$,
$B_\mu$ by

\begin{displaymath}
A_\mu = s_S A^{15}_\mu + \sqrt{1 - s_W^2 - s_S^2} B_\mu + s_W W_\mu^3 ,
\end{displaymath}

\noindent and two orthogonal to $A_\mu$ fields $Z_{1 \mu}$ and $Z_{2 \mu}$
can be written as

\begin{eqnarray}
Z_{1 \mu} & = & - t_W \big (s_S A^{15}_\mu + \sqrt{1 - s_W^2 - s_S^2}
B_\mu \big ) + c_W W_\mu^3 , \nonumber \\
Z_{2 \mu} & = & \big (\sqrt{1 - s_W^2 - s_S^2} A^{15}_\mu -
s_S B_\mu \big )/c_W , \nonumber
\end{eqnarray}

\noindent where $s_{W,S} = \sin \theta_{W,S}$, $c_W = \cos \theta_W$,
$t_W = \tan \theta_W$. The angles $\theta_W$ and $\theta_S$ of the weak
and strong mixings are defined as

\begin{eqnarray}
s_W^2 & = & \frac{\alpha(m)}{\alpha_W(m)} , \label{eq:sW} \\
s_S^2 & = & \frac{2}{3} \, \frac{\alpha(m)}{\alpha_{15}(m)} \, = \,
\frac{2}{3} \, \frac{\alpha(m)}{\alpha_S(m)} \, \left ( 1 +
\frac{\alpha_S(m)}{2 \pi} \, b \, \ln \frac{M_C}{m} \right ) , \label{eq:sS}
\end{eqnarray}

\noindent where $\alpha(m)$, $\alpha_W(m)$, $\alpha_S(m)$ are the
electromagnetic, weak and strong coupling constants at the scale $m$,
$M_C$ is the mass scale of the $SU_V(4)$-symmetry breaking,
$b = b_S - b_{15} = 11$, $b_S = 11 - \frac{2}{3} n_f$,
$b_{15} = - \frac{2}{3} n_f$, $n_f$ is the number of fermions with
masses below $M_C$.
The last equality in (\ref{eq:sS}) is obtained by the elimination of
the $SU_V(4)$ unified gauge coupling constant
$\alpha_4(M_C) = g_4^2 / 4 \pi$
from the one-loop approximation relations

\begin{equation}
\alpha_{S,15}(m) \, = \, \alpha_4(M_C)/\big ( 1 + \frac{\alpha_4(M_C)}
{2 \pi} \, b_{S,15} \, \ln \frac{M_C}{m} \big ) \label{eq:aS15}
\end{equation}

\noindent between $\alpha_4(M_C)$
and the $A^{15}$-interaction constant
$\alpha_{15}(m)$ and $\alpha_S(m)$.

The interaction of the gauge fields with the fermions can be
written as

\begin{eqnarray}
{\cal L}^\psi_{int} & = & \frac{g_4}{\sqrt 2} \big [ V^\alpha_\mu
\big ( \bar {Q'}_\alpha \gamma^\mu \ell' \big ) + h.c. \big ] +
\frac{g_2}{\sqrt 2} \big [ W^+_\mu \big ( \bar {\psi'}^L_{a A}
\gamma^\mu \big ( \tau^+ \big )_{a b} {\psi'}^L_{b A} \big ) + h.c.
\big ] \nonumber \\
& + & g_{st} G_\mu^j \big ( \bar Q \gamma^\mu t_j Q \big ) +
e A_\mu \big ( \bar \psi \gamma^\mu Q \psi \big ) + {\cal L}_{N C}.
\label{eq:Lint}
\end{eqnarray}

\noindent Here the first term describes the interaction
of leptoquarks with quarks and leptons by the constant $g_4$ related to
$\alpha_S(m)$ by (\ref{eq:aS15}). This interaction contains, in general,
the new generation mixing due to the matrices $K^{L,R}_a =
(A^{L,R}_{Q_a})^+ A^{L,R}_{\ell_a}$.
The second term in~(\ref{eq:Lint}) describes
the weak charged current interaction of $W^\pm$-bosons with quarks or
leptons by the constant $g_2$ related to the Fermi constant $G_F$
and $m_W$ in a usual way.
This interaction contains the well known Cabibbo-Kobayashi-Maskawa
mixing of the quarks due to the matrix
$C_Q = (A^L_{Q_1})^+ A^L_{Q_2}$
and analogous mixing in the lepton sector due to the lepton mixing matrix
$C_\ell = (A^L_{\ell_1})^+ A^L_{\ell_2}$.
The next two terms are the
QCD- and QED-interactions. The neutral current interaction
${\cal L}_{N C}$ can be written as

\begin{equation}
{\cal L}_{N C} = Z_\mu J^Z_{\mu} + Z'_\mu J^{Z'}_{\mu} , \label{eq:Lnc}
\end{equation}

\noindent where

\begin{eqnarray}
Z_\mu & = & Z_{1 \mu} \cos \theta_m + Z_{2 \mu} \sin \theta_m , \nonumber \\
{Z'}_\mu & = & - Z_{1 \mu} \sin \theta_m + Z_{2 \mu} \cos \theta_m , \nonumber
\end{eqnarray}

\noindent are the mass eigenstate fields and

\begin{eqnarray}
J_\mu^Z & = & J_\mu^{Z_1} \cos \theta_m + J_\mu^{Z_2} \sin \theta_m ,
\label{eq:JZ} \\
J_\mu^{Z'} & = & - J_\mu^{Z_1} \sin \theta_m + J_\mu^{Z_2} \cos \theta_m ,
\label{eq:JZ'} \\
J^{Z_1}_{\mu} & = & \frac{e}{s_W c_W} \, \big ( J^{3 L}_{\mu} -
   s_W^2 J^Q_\mu \big ), \label{eq:JZ1} \\
J^{Z_2}_\mu & = & \frac{e}{s_S c_W \sqrt{1 - s_W^2 - s_S^2}} \,
\Big [ c_W^2 \sqrt{\frac{2}{3}} J^{15}_{\mu} - s_S^2 \big (
J^Q_\mu - J^{3 L}_\mu \big ) \Big ]. \label{eq:JZ2}
\end{eqnarray}

\noindent with the currents
$J^Q_\mu = (\bar \psi_{p a A} \gamma_\mu Q_{a A} \psi_{p a A})$,
$J^{3 L}_\mu = \frac{1}{2} (\bar \psi_{p a A} \gamma_\mu (1 + \gamma_5)
(\tau_3/2)_{a a} \psi_{p a A})$,
$J^{15}_\mu = (\bar \psi_{p a A} \gamma_\mu (t_{15})_{A A} \psi_{p a A})$.
The $Z_1$-current (\ref{eq:JZ1}) is the usual neutral current of
the Standard Model, but the structure of the $Z_2$-current (\ref{eq:JZ2})
is specified by the model under consideration. The $Z-Z'$-mixing
angle $\theta_m$ is defined by the symmetry breaking mechanism of
the model and is found to be small.

The Higgs sector of the model is taken in the simplest way and consists
of the four multiplets (4, 1, 1), (1, 2, 1), (15, 2, 1), (15, 1, 0) of
$SU_V(4) \times SU_L(2) \times U_R(1)$-group with the vacuum expectation
values (VEV's) $\langle \phi_A^{(1)} \rangle = \delta_{A 4} \eta_1 /
\sqrt 2$, $\langle \phi_a^{(2)} \rangle = \delta_{a 2} \eta_2 / \sqrt 2$,
$\langle \phi_{a B}^{(3) A} \rangle = \delta_{a 2} ( t_{15} )^A_B \eta_3$
and $\langle \phi_B^{(4) A} \rangle = ( t_{15} )^A_B \eta_4$ respectively.
After breaking the symmetry in such a way the masses of quarks
and leptons are defined
by VEV's $\eta_2$, $\eta_3$ and by Yukawa coupling constants and may be
arbitrary just as they are in the Standard Model; the photon and the gluons
are still massless but all the other gauge fields acquire the masses.
The VEV $\eta_4$ contributes only to the masses of leptoquarks,
and hence the leptoquarks may be heavy enough irrespective of the other
gauge particles.

For the masses of the $W$-, $Z$- and $Z'$-bosons
the model predicts the mass relation

\begin{equation}
\big ( \mu^2 - \rho_0 \big ) \big ( \rho_0 - 1 \big ) = \rho^2_0 \sigma^2 ,
\label{eq:mr}
\end{equation}

\noindent where $\mu \equiv m_{Z'} / m_Z$, $\rho_0 \equiv m_W^2 / m_Z^2
c_W^2$ and

\begin{equation}
\sigma = \frac{s_W s_S}{\sqrt{1 - s_W^2 - s_S^2}}. \label{eq:s}
\end{equation}

\noindent Simultaneously the model gives
for the $Z$ - $Z'$ mixing angle $\theta_m$
the expression

\begin{equation}
\sin \theta_m = \Big [ 1 + \Big ( \frac{\rho_0 \sigma}{\rho_0 - 1} \Big )^2
\Big ]^{-1/2}  \label{eq:sint}
\end{equation}

\noindent For $\theta_m \ll 1$ and $\rho_0 \simeq 1$
we also obtain from (\ref{eq:mr}), (\ref{eq:sint}) that $\theta_m
\simeq \sigma \, m_Z^2 / m_{Z'}^2$.

It should be noted that all the coupling constants in (\ref{eq:Lint}),
(\ref{eq:Lnc}) may be
expressed by means of $s_W$, $s_S$ in terms of the known coupling constants
$\alpha(m)$, $\alpha_W(m)$, $\alpha_S(m)$
at the scale $m = m_Z$ and the $SU_V(4)$ unification mass scale $M_C$
which enters $s_S$ and $\alpha_4(M_C)$ accoding to (\ref{eq:sS}),
(\ref{eq:aS15}) and is found to be the only unknown parameter of the
Lagrangian (\ref{eq:Lint}), (\ref{eq:Lnc}). Of course,
to calculate the neutral current processes described by (\ref{eq:Lnc})
it is necessary to know $m_{Z'}$, which actually results in two
unknown parameterts $M_C$ and $m_{Z'}$.

The mass relation (\ref{eq:mr}) gives the limit on the $Z'$-mass.
Using the experimental values~\cite{PDG}
of $G_F$, $m_W$ and $\alpha(m_Z)$
we have $s_W^2 = 0.2298 \pm 0.0014$. Then
taking the most stringent limit $M_C \ge 10^5 \div 10^6 \, GeV$
resulting from~\cite{PDG}
$Br (K^0_L \rightarrow \mu e) < 0.94 \cdot 10^{-10}$
into account and using the experimental values~\cite{PDG}
$\alpha_s(m_Z) = 0.1134 \pm 0.0035$
we evaluate $s_S^2$ and $\sigma$ from~(\ref{eq:sS}) and~(\ref{eq:s}) for
$M_C = 10^{6} \div 10^{14} \, GeV$ (see Table 1).
\begin{table}[tb]
\hspace{0.2\hsize}\parbox[b]{0.6\hsize}{\caption{The strong mixing angle
$\sin^2 \theta_S$ and the parameter
$\sigma$ depending on the mass scale $M_C$ in the MQLS-model.}}
\begin{center}
\begin{tabular}{ccc}\hline
$M_C$, $GeV$ & $\sin^2 \theta_S$ & $\sigma$   \\ \hline
$10^6$       &  0.130            &  0.216     \\
$10^{10}$    &  0.213            &  0.297     \\
$10^{14}$    &  0.297            &  0.380     \\ \hline
\end{tabular}
\end{center}
\end{table}
For these values of the $\sigma$ the relation $m_{Z'}/m_Z$
and $\sin \theta_m$ as functions of the $\Delta \rho_0 \equiv \rho_0 - 1$
are presented on Fig.1.
\begin{figure}[htb]
\unitlength=1mm
\begin{picture}(150,100)
\put(15,0){\parbox[b]{0.8\hsize}{\caption{Mass relation $m_{Z'}/m_Z$
         and $\sin \theta_m$ as functions
         of the $\Delta \rho_0$ in MQLS-model: a)~$M_C=10^6 \, GeV$,
         b)~$M_C=10^{10} \, GeV$, c)~$M_C=10^{14} \, GeV$.}}}
\end{picture}
\end{figure}
Extracting the allowed values of $\Delta \rho_0$
from the experimental value \cite{PDG}

\begin{displaymath}
\rho_{eff} = \rho_0 \Big ( 1 + \frac{3}{8} \> \frac{G_F m_t^2}{\sqrt 2 \pi^2}
\Big ) = 1.007 \pm 0.002
\end{displaymath}

\noindent we obtain $\Delta \rho_0 = (4~\pm~2)~\cdot~10^{-3}$,
$(2~\pm~2)~\cdot~10^{-3}$, $(0~\pm~2)~\cdot~10^{-3}$ for
$m_t = 100 \, GeV, \; 125 \,
GeV$ and $150 \, GeV$ respectively. Then we see from Fig.1 that the
$Z'$-boson may be rather light.
Thus for $M_C = 10^6 \, GeV$
and $m_t = 125 \, GeV$ we get the limits  $m_{Z'} > 4\, m_Z$
on $Z'$-mass and $\theta_m < 0.01$ on the $Z$-$Z'$-mixing
angle. This upper limit on $\theta_m$ is compatible with those obtained
in the extended gauge models \cite{LL,R,A}.

Using the structure (\ref{eq:Lnc}) - (\ref{eq:JZ2}) of the neutral current
interaction we have calculated in the tree approximation the cross
sections
$\sigma_{\bar f f} = \sigma (e^+ e^- \rightarrow \gamma, Z, Z'
\rightarrow \bar f f)$.
The leptonic cross section $\sigma_{\bar \ell \ell}$
is found to be less than the one predicted by the SM.
This effect is due to
the destructive $\gamma$-$Z'$-interference~\cite{PSch}. The magnitude
of this deviation depends on the MQLS-model parameters $m_{Z'}$ and $M_C$
and can be used for the derivations of the limits on $m_{Z'}$ and $M_C$
from the experimental data. Calculating the relative deviations
$\delta_{\bar \ell \ell} = ( \sigma_{\bar \ell \ell} -
\sigma_{\bar \ell \ell}^{SM}) / \sigma_{\bar \ell \ell}^{SM}$
at $\sqrt s = 60 \, GeV$ and using the current values of the leptonic
cross section measured by AMY-, VENUS- and TOPAZ- groups
at TRISTAN~\cite{S}, we have obtained the limits on $Z'$-mass
in dependence on $M_C$ (or $\sin^2 \theta_S$). As an example, such
limits on $m_{Z'}$ extracted from $\sigma_{\mu^+ \mu^-}$ and
corresponding to the experimental error of the one standard deviation
are presented on Fig.~2.
\begin{figure}[htb]
\unitlength=1mm
\begin{picture}(150,140)
\put(8,0){\parbox[b]{0.9\hsize}{\caption{The limits on $Z'$-mass in
MQLS-model extracted from the muonic
cross sections $\sigma_{\mu^+ \mu^-}$ measured by AMY-, VENUS- and TOPAZ-
groups at TRISTAN$^7$ (68 \% CL).}}}
\end{picture}
\end{figure}
We see from Fig.~2 that within the experimental
error of one standard deviation the current
$\sigma_{\mu^+ \mu^-}$
data~\cite{S} give the lower limit on $m_{Z'}$ about a few hundreds $GeV$
depending on $M_C$. As concerns the upper limits on $m_{Z'}$ extracted
from $\sigma_{\mu^+ \mu^-}$ VENUS- and TOPAZ- data, they should be
regarded as preliminary because these limits depend crucialy on the
experimental errors and can be taken into account more seriously only
at high experimental accuracy.

The hadronic cross section $\sigma_h = \sum_q \sigma_{\bar q q}$
is found to be somewhat more than that predicted by the SM
but this deviation is smaller than that in the leptonic case.
Hence, the measurements of the
leptonic cross sections are more favourable for the search
for the possible manifestations of the extra $Z'$-boson than
the measurements of the hadronic cross sections.

We have calculated also the forward-backward asymmetry $A_{FB}$
of $e^+ e^- \rightarrow \gamma, Z, Z' \rightarrow \bar \ell \ell$
reactions and have analysed the limits on $Z'$- mass from $A_{FB}$-
asymmetry measured at TRISTAN~\cite{S}. These limits turned out
to be weaker than those extracted from cross sections
$\sigma_{\mu^+ \mu^-}$.

\begin{figure}[htb]
\unitlength=1mm
\begin{picture}(150,140)
\put(8,0){\parbox[b]{0.9\hsize}{\caption{The possible lower limits on
$Z'$-mass in MQLS-model
from future measurements of the deviations $\delta_{\bar \ell \ell}$
of the leptonic cross sections $\sigma_{\bar \ell \ell}$ from the
SM predictions at LEP200.}}}
\end{picture}
\end{figure}
The evaluation of leptonic cross sections $\sigma_{\bar \ell \ell}$
at $\sqrt s = 200 \, GeV$ shows that their deviations
$\delta_{\bar \ell \ell}$ from the SM predictions are significantly
larger at such energies and, hence, the measurements of these deviations
at LEP200 will allow either to observe the manifestation of $Z'$-boson
originated from the four-color quark-lepton symmetry or
to obtain the most stringent limits on the MQLS-model
parameters~$m_{Z'}$ and~$M_C$.
On Fig.~3 we show the lower limits on $m_{Z'}$ in dependence on $M_C$
(or $\sin^2 \theta_S$) which can be achieved at LEP200 by measurements
of the deviations $\delta_{\bar \ell \ell}$ of leptonic cross sections
from the SM predictions to an accuracy of 20\%, 10\% and 5\% if
the results of these measurements will be negative. It can be seen from
Fig.~3 that, in particular, the lower limit on $m_{Z'}$ for
$M_C \sim 10^6 \, GeV$ and at 5\% accuracy of $\delta_{\bar \ell \ell}$
measurements can be raised up to about $1.2 \, TeV$.

In summary we can conclude that the minimal extension of the Standard
Model containing the four-color quark-lepton symmetry discussed here
leads, in particular, to the existence of an additional $Z'$- boson
originated from this symmetry. The limits on its mass mass $m_{Z'}$
obtained here from the mass relation and from the current TRISTAN data
on $\sigma_{\bar \ell \ell}$, $\sigma_h$ and $A_{FB}$ show that such
$Z'$- boson may be rather light ($m_{Z'}$ is larger a few hundred $GeV$)
and may be interesting for experimental search at LEP200 and
future colliders.

\bigskip

\noindent {\bf Acknowledgments}

\bigskip

\noindent The author is grateful to V.A.~Rubakov for the helpful discussion
of the results and to the Organizing Committee of the International Seminar
``Quarks-94'' for the possibility to participate in this Seminar.
The work was supported by the Russian Foundation for Fundamental
Research (Grant No. 93-02-14414).


\begin{thebibliography}{7}
\bibitem{PSm}
   J.C.~Pati and A.~Salam, {\it Phys.~Rev.} {\bf D10}~(1974)~275.
\bibitem{PDG}
   Particle Data Group, K.~Hikasa et~al., {\it Phys.~Rev.} {\bf D45}~(1992)
   N11, part~2.
\bibitem{LL}
   P.~Langacker and M.~Luo, {\it Phys.~Rev.} {\bf D45}~(1992)~278.
\bibitem{R}
   S.~Riemann, {\it DESY preprint} DESY 92-143 (1992).
\bibitem{A}
   G.~Altarelli et al., {\it CERN preprint} CERN-TH.6947/93 (1993).
\bibitem{PSch}
   A.A.~Pankov and J.S.~Satsunkevich, {\it Yad.~Fiz.} {\bf 47}~(1987)~1333.
\bibitem{S}
   M.~Sakuda, {\it KEK preprint} 93-124 (1993).
\end{thebibliography}
\end{document}